\documentclass[letter]{aa}

\pdfoutput=1

\usepackage{graphicx}
\usepackage{amssymb}
\usepackage{epstopdf}
\usepackage{isolatin1} 		
\usepackage{gensymb} 		
\usepackage{url}
\usepackage{hyperref}		
\usepackage{natbib}
\bibpunct{(}{)}{;}{a}{}{,} 		
\usepackage{txfonts} 		

\hypersetup{
    colorlinks=false,
    urlcolor=blue,
    bookmarks,
    bookmarksnumbered,
    pdftitle={Hard X-ray emission from Eta Carinae},
    pdfauthor={Leyder J.-C., Walter R. and Rauw G.},
    pdfsubject={INTEGRAL detects hard X-ray emission from Eta Carinae},
    pdfkeywords={gamma rays: observations  - X-rays: binaries - X-rays: individuals:  Eta Carinae - X-rays: individuals: 1E~1048.1-5937 - X-rays: individuals: IGR~J10447-6027},
    pdfcenterwindow=true
}

\begin{document}

\title{Hard X-ray emission from $\eta$~Carinae}

\author{
J.-C. Leyder \inst{1,2} \fnmsep\thanks{FNRS Research Fellow} \and
R. Walter \inst{2,3} \and
G. Rauw \inst{1}  \fnmsep\thanks{FNRS Research Associate} 
}

\offprints{J.-C. Leyder}

\institute{
Institut d'Astrophysique et de Géophysique, Université de Liège, Allée du 6-Août 17, Bâtiment B5c, B--4000 Liège, Belgium \\ \email{leyder@astro.ulg.ac.be} \and
\textit{INTEGRAL} Science Data Centre, Université de Genève, Chemin d'Écogia 16, CH--1290 Versoix, Switzerland \and
Observatoire de Genève, Université de Genève, Chemin des Maillettes 51, CH--1290 Sauverny, Switzerland
}

\date{Received 2 November 2007 / Accepted 16 November 2007}

\abstract
{
If relativistic particle acceleration takes place in colliding-wind binaries, hard X-rays and $\gamma$-rays are expected through inverse Compton emission, but to date these have never been unambiguously detected.
}
{
To detect this emission, observations of \object{$\eta$~Carinae} were performed with \textsc{INTEGRAL}, leveraging its high spatial resolution.
}
{
Deep hard X-ray images of the region of \object{$\eta$~Car} were constructed in several energy bands.
}
{
The hard X-ray emission previously detected by \textsc{BeppoSax} around \object{$\eta$~Car} originates from at least 3 different point sources. The emission of \object{$\eta$~Car} itself can be isolated for the first time, and its spectrum unambiguously analyzed. The X-ray emission of \object{$\eta$~Car} in the 22--100~keV energy range is very hard ($\Gamma \simeq 1 \pm 0.4$) and its luminosity is $7 \times 10^{33}$~erg\,s$^{-1}$.
}
{
The observed emission is in agreement with the predictions of inverse Compton models, and corresponds to about 0.1\% of the energy available in the wind collision.
\object{$\eta$~Car} is expected to be detected in the GeV energy range.
}




\maketitle

\section{Introduction}
\label{sec:Introduction}
\object{$\eta$~Carinae} is one of the most peculiar objects in our Galaxy \citep[see][]{Davidson+97}. Once the second brightest object in the sky (during its eruption in 1843), it decreased to a V magnitude $m_{\mathrm{V}}$ of 8 by the end of the XIXth century, before slowly and irregularly increasing again, up to its current value of $m_{\mathrm{V}} \sim 5$ \citep[see e.g.][]{Viotti-95}.
The large quantities of matter that were ejected during these dramatic luminosity variations are now forming an extended nebula (the so-called \textit{homunculus}), while \object{$\eta$~Car} is still ejecting matter through energetic stellar winds. Observations lead to an estimated mass-loss rate of $10^{-4}$--$10^{-3}$~$M_{\sun}$\,yr$^{-1}$ \citep{Andriesse+78, Hillier+01, Pittard+02-EtaCar, vanBoekel+03}.

Optical spectra of \object{$\eta$~Car} reveal long periods of ``high spectroscopic state'' characterized by an emission line spectrum, followed by shorter periods of ``low spectroscopic state'' --- also called ``spectroscopic events'' --- typically lasting a few months, during which the high-excitation emission lines fade away \citep{Rodgers+67, Viotti-69, Zanella+84, Altamore+94}.

A period of $\sim$2023~days (5.53~yr) was inferred from periodic changes in the optical \citep{Damineli-96, Damineli+00} and IR \citep{Whitelock+94, Whitelock+04} domains.
There is now strong evidence that $\eta$~Car is a binary system; some examples are the radial velocity variations occurring with the period of 5.53~yr, or the ``spectroscopic event'' (believed to occur near periastron passage).

Many X-ray observations have also been performed, which have extended our understanding of the physical nature of \object{$\eta$~Car}.
The structured X-ray emission can be divided into two components, as shown by the \textit{Einstein} satellite \citep{Chlebowski+84, Rebecchi+01, Viotti+02-JApA}, and further confirmed by \textit{Chandra} observations \citep{Seward+01}:\\
-- a soft ($kT_{\mathrm{SX}} \sim 0.5$~keV) thermal X-ray component ($\eta$\,SX), which is spatially extended (up to about 20\arcsec) and in-homogenous, dominates the spectrum mostly below 1.5~keV, and is probably associated to the interaction of the stellar wind with the interstellar matter;\\
-- a hard ($kT_{\mathrm{HX}} \sim 4.7$~keV) thermal X-ray component ($\eta$\,HX), which is point-like, centered on the stellar system, dominates the spectrum in the 2--10~keV range, and is likely linked to the wind collision between the two massive stars that form the binary system.

The X-ray light curves also exhibit strong variations, with the same periodicity of $2024\pm2$~days \citep{Ishibashi+99, Corcoran-05}. The X-ray spectrum suggests a colliding-wind binary (CWB) scenario \citep{Ishibashi+99, Pittard+02-EtaCar, Corcoran-05}: the dense stellar wind coming from the massive luminous blue variable (LBV) primary star ($\eta$~Car~A) collides with the higher-velocity, lower-density wind from the hotter and luminous, probably late-type nitrogen-rich O or WR type \citep{Iping+05}, star companion ($\eta$~Car~B) in a highly eccentric orbit.

Finally, at hard X-rays, \textit{BeppoSAX} detected \object{$\eta$~Car} only once exhibiting a hard X-ray tail (\textit{i.e.} at a flux higher than an extrapolation of the hard thermal X-ray component). In this Letter, new high-resolution hard X-ray observations performed by \textit{INTEGRAL} are presented, which allow to clearly detect the emission from \object{$\eta$~Car} for the first time.
Such a hard X-ray detection is important, as it proves that non-thermal particle acceleration is at work in the wind collision, and as it suggests likely gamma-ray emission in the MeV and GeV energy ranges.

\section{High-energy observations and data analysis}
\label{sec:Observations}

\subsection{Previous \textit{BeppoSAX} observations}
\label{subsec:BeppoSAX}
The \textit{BeppoSAX} satellite carried out 4 observations of \object{$\eta$~Car} between 1996 and 2000, covering its 5.53~yr cycle ($\Phi = 1$ corresponds to the 1998 minimum):\\
-- December 1996 ($\Phi = 0.83$), December 1999 ($\Phi = 1.37$), and June 2000 ($\Phi = 1.46$) in the high spectroscopic state;\\
-- March 1998 ($\Phi = 1.05$) in the low spectroscopic state.

The first hard X-ray detection at energies between 10 to 20~keV in the vicinity of \object{$\eta$~Car} was obtained with the phoswich detector system (PDS) instrument on-board \textit{BeppoSAX} in December 1996, when the 13--20~keV flux ($0.15-0.17$~count\,s$^{-1}$) was clearly in excess of the 4.7~keV thermal fit found from the 2--10~keV energy range \citep{Rebecchi+01, Viotti+02-A&A}. This flux excess is found again in the other two observations in the high spectroscopic state (December 1999 and June 2000), but not in the low spectroscopic state observation (March 1998) where only an upper limit on the 13--20~keV flux could be extracted \citep{Viotti+04}.

Moreover, the June 2000 observation was longer (nominal exposure time of 100~ks), and revealed a high-energy tail extending up to 50~keV.
This emission, detected with the non-imaging PDS instrument, was attributed to \object{$\eta$~Car}, based mostly on the fact that \object{$\eta$~Car} is the strongest and hardest source seen in the simultaneous medium energy concentrator spectrometer (MECS) 1.5--10~keV imaging observations \citep{Viotti+02-A&A, Viotti+04}. 
However, the \textit{INTEGRAL} observations presented in Sect.~\ref{subsec:INTEGRAL} indicate the presence of at least two additional hard X-ray sources in the PDS field of view (FOV; see Fig.~\ref{fig:ISGRI-22-100}), but which were not within the MECS FOV. Therefore, a significant fraction of the hard X-ray flux detected by PDS originates from sources nearby \object{$\eta$~Car}, and could affect the PDS spectral analysis performed by \citet{Viotti+04}.

\subsection{\textit{INTEGRAL} observations}
\label{subsec:INTEGRAL}
The ESA \textit{INTEGRAL} $\gamma$-ray satellite carries (in addition to an optical monitoring camera) 3 co-aligned instruments dedicated to the observation of the high-energy sky, from 3~keV up to 10~MeV \citep{WInkler+03}. The \textit{INTEGRAL} soft gamma-ray imager (ISGRI; \citealt{Lebrun+03}), the most sensitive detector between 15 and 200~keV, offers the first unambiguous detection of \object{$\eta$~Car} at hard X-rays.

All available public \textit{INTEGRAL} data located within $10\degree$ of \object{$\eta$~Car} were selected, resulting in 1131 pointings, for a total elapsed observing time of 3.3~Ms and a deadtime-corrected good exposure of 2.3~Ms. There are 3 major periods during which the source was frequently observed, listed in Table~\ref{tab:INTEGRAL-data}. Together, they represent about 85\% of the data used; the remaining observations come from Galactic plane scans, and are therefore well spaced over time. The effective exposure time (corresponding to an on-axis observation) amounts to 1.1~Ms (in the 22--35~keV energy range). ISGRI pointing sky images were produced using OSA\footnote{The Offline Scientific Analysis (OSA) software is available from the ISDC website~: \url{http://isdc.unige.ch}}, version 6.0, with standard parameters. Good time intervals were built using a strong constraint on the attitude stability ($\Delta<3\arcsec$). The image cleaning step used an input catalogue of 24 sources with fixed source positions, and was applied independently of the source strength (thus allowing for negative source models) in order to avoid introducing any bias in the process.

A broad band (22--100~keV) mosaic image of the field was produced, along with narrower bands in 3 energy ranges~: 22--35~keV, 35--50~keV, 50--100~keV.
The most external parts of each individual image were skipped, as they are more noisy but do not add much signal. The final mosaic images were built in equatorial coordinates with a tangential projection, using an over-sampling factor of two with respect to the individual sky images. The photometric integrity and accurate astrometry are obtained by calculating the intersection between input and output pixels, and weighting the count rates with the overlapping area.
\begin{table}[htdp]
\caption{Main \textit{INTEGRAL} public data available for \object{$\eta$~Car}.}
\label{tab:INTEGRAL-data}
\centering
\begin{tabular}{ccccc} \hline \hline
Period & Rev. & Time [MJD] & Phase $\Phi$ of \object{$\eta$~Car} \\ \hline 
1 & 76--88 & 52\,787--52\,827 & 1.99--2.01 \\
2 & 192--209 & 53\,134--53\,188 & 2.16--2.19\\
3 & 322--330 & 53\,523--53\,550 & 2.35--2.37 \\ \hline
\end{tabular}
\end{table}

\begin{figure}[htbp]
\centering
\includegraphics[width=0.8\columnwidth]{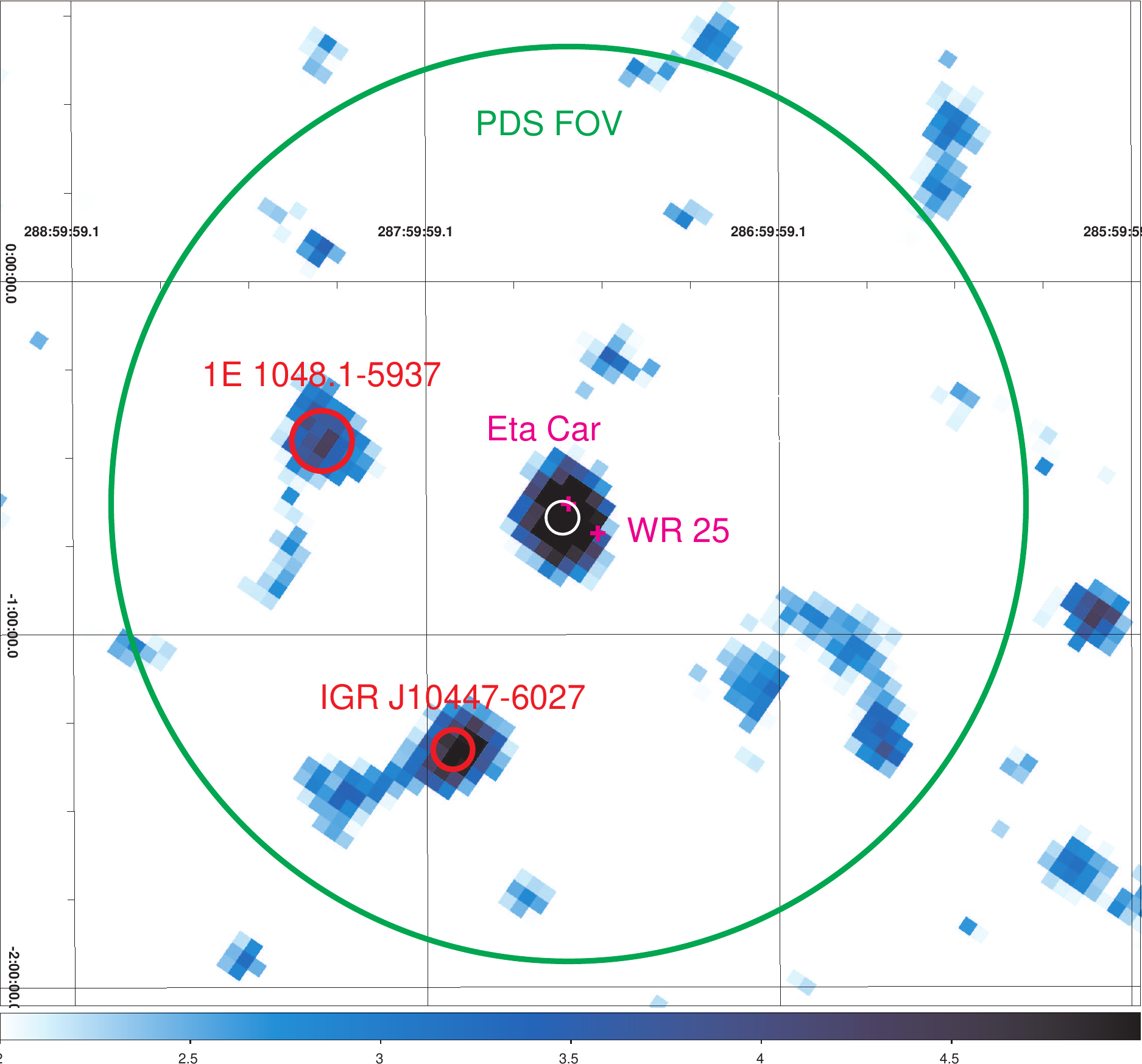}
\caption{ISGRI significance image (22--100~keV; effective exposure time of 1.1~Ms; significance goes from 2$\sigma$ to 5$\sigma$) around \object{$\eta$~Car}, showing the positions of  \object{$\eta$~Car} and \object{WR~25} (in magenta). The enclosing green circle symbolizes the PDS FOV. The white and red circles represent the best-fit positions listed in Table~\ref{tab:INTEGRAL-sources}.}
\label{fig:ISGRI-22-100}
\end{figure}

The ISGRI 22--100~keV image is shown in Fig.~\ref{fig:ISGRI-22-100}. Several flux excesses are clearly detected; Table~\ref{tab:INTEGRAL-sources} summarizes their best-fit position, error circle, intensity and significance. Besides \object{$\eta$~Car}, one source has an error circle that includes the anomalous X-ray pulsar (AXP) \object{1E 1048.1-5937}, and another one (named \object{IGR~J10447-6027}) is coincident with the South of the giant dust pillar nebula of the Carina region, and in particular with a massive young stellar object (YSO; \object{IRAS 10423-6011}), probably corresponding to an embedded B0 star \citep{Rathborne+04}. The emission observed by \textit{INTEGRAL} for the latter source could be interpreted either as evidence for a new high-mass X-ray binary, or as a signature of accretion and/or particle acceleration in the YSO. A significant fraction of the flux detected with PDS is likely to originate not only from \object{$\eta$~Car} but also from these other two sources, in proportions that depend on both the energy range and the time.

\begin{table*}[htdp]
\caption{Sources detected within the PDS FOV around \object{$\eta$~Car} (the flux and significance values were extracted from the 22--100~keV energy band, with the best-fit position, and with the PSF size left free to vary; the error circle corresponds to a 90\% probability).}
\label{tab:INTEGRAL-sources}
\centering
\begin{tabular}{ccccc} \hline \hline
Source & Intensity [cnt\,s$^{-1}$] & Position (J2000) & Error circle radius [\arcmin] & Significance\\ \hline 
$\eta$~Car & $0.16 \pm 0.02$ & RA~= $10^\mathrm{h}45^\mathrm{m}02$, Dec~= $-59\degree43\arcmin38$ & 2.8 & 7.9 \\
1E 1048.1-5937 & $0.09 \pm 0.02$ & RA~= $10^\mathrm{h}50^\mathrm{m}38$, Dec~= $-59\degree50\arcmin40$ & 5.1 & 4.5\\
IGR~J10447-6027 & $0.12 \pm 0.02$ & RA~= $10^\mathrm{h}44^\mathrm{m}47$, Dec~= $-60\degree27\arcmin15$ & 3.4 & 5.8 \\ \hline
\end{tabular}
\end{table*}

The most significant source is coincident with the most powerful X-ray source in the field (\object{$\eta$~Car}), although it should be noted that another close X-ray source (\object{WR~25}) lies just outside of the error box.
However, the shape of the point-spread function (PSF) in the best fit (5.4\arcmin\ by 6.3\arcmin) is consistent with a point source. Moreover, the positions extracted from the images (in all individual energy bands, as well as in the 22--100~keV image) indicate that only \object{$\eta$~Car} is consistently inside the error box, while \object{WR~25} always lies close, but outside. The angular separation between \object{$\eta$~Car} and \object{WR~25} is larger than 7\arcmin, as opposed to the typical error box of 3\arcmin.
Therefore, the hard X-ray emission observed with \textit{INTEGRAL} can be associated with \object{$\eta$~Car}.

Based on the 22--100~keV images, Table~\ref{tab:INTEGRAL-results} lists (for the three major observing periods, as well as for the whole data set) the effective exposure, the count rate and the detection significance of \object{$\eta$~Car} (or the $3\sigma$-upper limit in the case of the first period where the object is not detected).
As expected from the highly uneven effective exposure durations of the 3 different major observing periods, the source is not detected during the first period, well detected during the second one, and slightly below the detection level during the third period. 

\begin{table}[htdp]
\caption{\textit{INTEGRAL} observations of \object{$\eta$~Car} (22--100~keV fluxes extracted by fixing the position of  \object{$\eta$~Car} and the PSF size to 6\arcmin).}
\label{tab:INTEGRAL-results}
\centering
\begin{tabular}{cccc} \hline \hline
Period & Eff. exp. [ks] & Count rate [cnt\,s$^{-1}$] & Significance\\ \hline 
1 & 122 & $<0.19$ & --- \\
2 & 717 & $0.16\pm0.03$ & 6.2 \\
3 & 180 & $0.18\pm0.05$ & 3.3 \\ \hline
All data & 1113 & $0.15\pm0.02$ & 7.3 \\ \hline
\end{tabular}
\end{table}

The average source flux observed by ISGRI, extracted from this image by fixing the position of the object and by assuming a PSF of 6\arcmin, is $0.15 \pm 0.02$~count\,s$^{-1}$ in the 22--100~keV energy range. This corresponds to a flux of $1.11 \times 10^{-11}$~erg\,cm$^{-2}$\,s$^{-1}$ when assuming a photon index of 1. Based on a distance of 2.3~kpc \citep{Smith-06}, the hard X-ray luminosity is $7 \times 10^{33}$~erg\,s$^{-1}$ (22--100~keV).

\begin{figure}[htbp]
\centering
\includegraphics[width=0.6\columnwidth, angle=90]{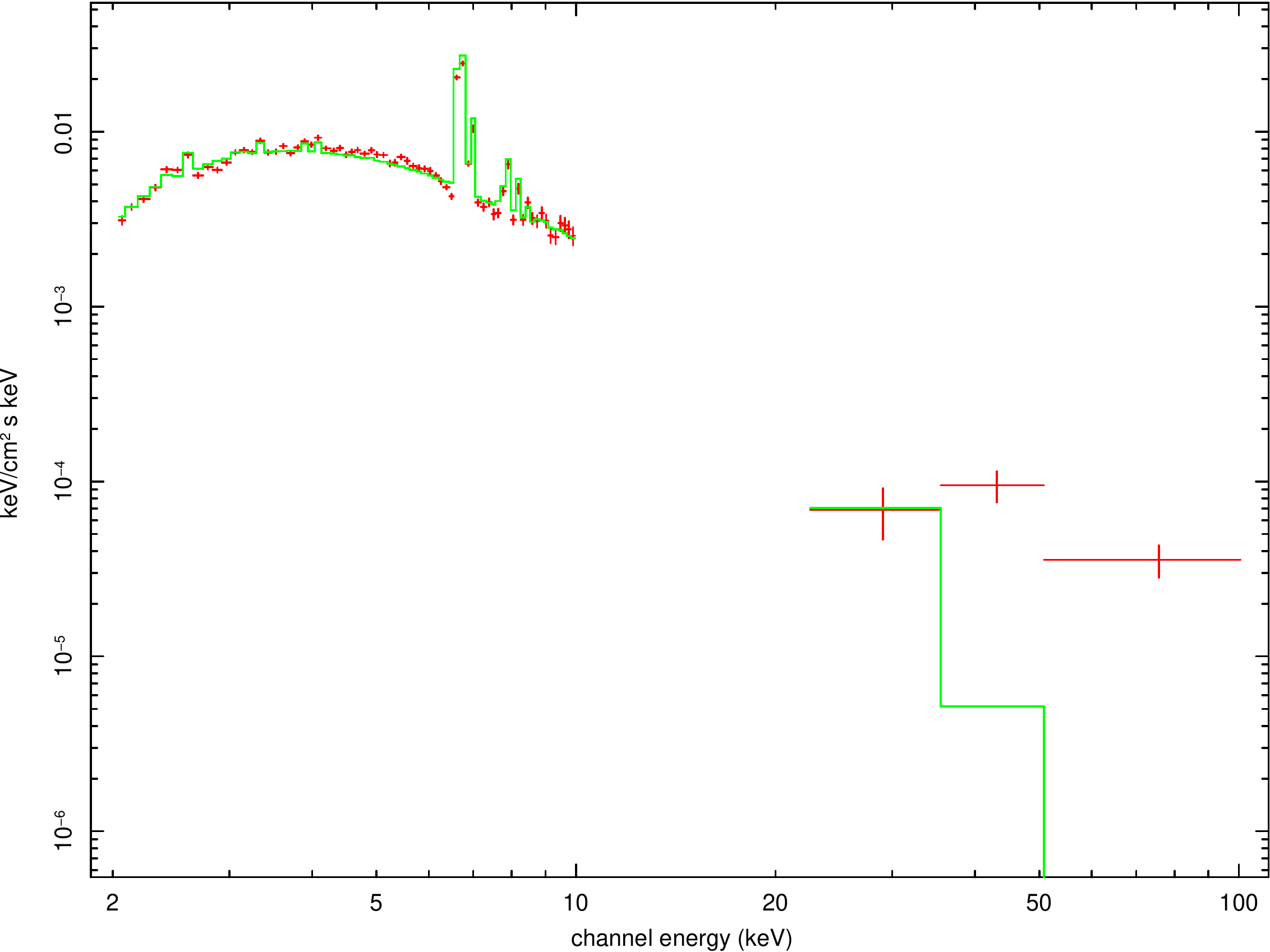}
\caption{Unfolded spectrum of \object{$\eta$~Car}, with MECS (2--10~keV) and ISGRI (22--100~keV) data, both shown with red crosses. The data are fitted with a \texttt{wabs*mekal} model (in green; $kT \simeq 5.1$~keV).}
\label{fig:MECS-ISGRI-euf}
\end{figure}

The ISGRI unfolded spectrum (extracted under the same conditions) is shown in Fig.~\ref{fig:MECS-ISGRI-euf}, together with the archival MECS spectrum from June 2000. They are fitted with a \texttt{wabs*mekal} model ($kT \simeq 5.1$~keV and $N_{H} \simeq 4.3 \times 10^{22}$~cm$^{-2}$, both in agreement with \citealt{Viotti+04}). This \texttt{mekal} model provides an excellent fit to the MECS data, and to the first ISGRI bin (\textit{i.e.} up to 30~keV), but is unable to reproduce the spectrum at higher energy. Given the limited number of points in the ISGRI data, the spectral shape of the high-energy emission is poorly constrained. A simple \texttt{powerlaw} model fit to the hard X-ray data gives a photon index $\Gamma$ around $1 \pm 0.4$ (with harder values reached when used in combination with the \texttt{mekal} component to fit the low-energy part of the ISGRI spectrum).

The photon index is much harder than found by \citet{Viotti+04}, but the data are not simultaneous, and the 2 other sources within the PDS FOV could have been responsible for a significant contribution.
The ISGRI spectrum extends at least up to 100~keV, while the PDS X-ray tail does not go beyond 50~keV. At energies between 20 and 50~keV, the average spectrum obtained from ISGRI observations is much weaker than the PDS spectrum from June 2000, while at energies above 50~keV, the ISGRI flux is stronger. Hence, the ISGRI photon index is much harder than an extension of the hard X-ray thermal component observed below 10~keV.

Based on ISGRI observations listed in Table~\ref{tab:INTEGRAL-results}, the tiny difference in observing time between periods 1 and 3 does not justify the total absence of source detection in period 1. Therefore, it seems that some variability is observed, in agreement with that reported by \citet{Viotti+04} following the high and low spectroscopic states. However, given the presence of 3 sources within the PDS FOV, it is unclear whether this variation can be trusted in the \textit{BeppoSAX} data. In particular, the flux observed between 20 and 30~keV with ISGRI is roughly 10~times lower than the flux observed with PDS, and this difference could perhaps be attributed to the 2 additional sources. It should also be noted that the AXP is variable, and exhibited a new X-ray burst detected by \textit{Swift} in March 2007, with a flux up to $6.3 \times 10^{-11}$~erg\,s$^{-1}$\,cm$^{-2}$ (in the 1--10~keV energy range; \citealt{Campana+07}); this is 8~times brighter than observed in the averaged ISGRI mosaics.

\section{Discussion}
\label{sec:Discussion}
The high-energy flux excess observed at the position of \object{$\eta$~Car} reveals unambiguously for the first time the presence of non-thermal emission at hard X-rays in a CWB.
Although possibilities, such as an unknown background object or a highly obscured super-giant X-ray binary, may be envisioned, it is unlikely that such an object would not have been detected in X-rays, especially in the frequently observed Carina region.

Inverse Compton (IC) scattering of low-frequency photons by high-energy electrons accelerated in the wind collision zone of CWBs \citep[see e.g.][]{Benaglia+03} is proposed as the emission mechanism responsible for the hard X-ray detection.
As explained in Sect.~\ref{sec:Introduction}, \object{$\eta$~Car} is an eccentric CWB, thus with a long orbital period and a large orbital separation.
This implies that the winds from the two members of the binary system reach their terminal velocity before colliding\footnote{Given the high eccentricity of \object{$\eta$~Car} ($e \sim 0.9$), the separation at periastron could mean the primary wind does not reach its terminal velocity. However, as shown by Table~\ref{tab:INTEGRAL-results}, the observations are dominated by the phases when \object{$\eta$~Car} is far from periastron.}, leading to strong shocks with high temperatures. 
This also means that the density of UV stellar photons is relatively low in the shock zone between the winds\footnote{Unless one of the two stars has a very weak wind, thus implying that the collision region would be displaced very close towards that star. This seems however unlikely, given what is already known about their parameters from the analysis of the X-ray lightcurve of \object{$\eta$~Car} (see below).}, as opposed to CWBs with short orbital periods. The question \textit{a priori} is therefore whether the level of IC emission is sufficient to allow a detection against the thermal emission from the shocked winds.

The total power contained in stellar wind interactions can be evaluated as: $L = 0.5 \Xi \dot{M} v^{2}$ \citep{Pittard+02-Wind}.
For \object{$\eta$~Car}, the parameters from \citet{Pittard+02-EtaCar} can be adopted~: $\dot{M_1} = 2.5 \times 10^{-4}$~$M_{\odot}$/yr, $v_{\infty,1} = 500$~km\,s$^{-1}$, $\dot{M_2} = 1 \times 10^{-5}$~$M_{\odot}$/yr, and $v_{\infty, 2} = 3000$~km\,s$^{-1}$; corresponding to $\eta = (\dot{M_2} v_{\infty,2})/(\dot{M_1} v_{\infty,1}) = 0.2$ (hence $\Xi_{1} \simeq 0.05$ and $\Xi_{2} \simeq 0.35$). These values yield a total power $L_{1}+L_{2}$ of $\sim 10^{37}$~erg\,s$^{-1}$, which is potentially available for thermal and non-thermal emission. The luminosity detected by ISGRI represents only 0.1\% of the total power involved in the stellar wind interactions.

\object{$\eta$~Car} shares some properties with \object{WR~140} (e.g. long period, high eccentricity). The non thermal radio detection of \object{WR~140} allowed \citet{Pittard+06} to estimate the expected (but not yet observed) hard X-ray IC emission as 0.5\% of the kinetic power involved in the wind collision, close to the value observed in \object{$\eta$~Car}. However, \object{WR~140} is not detected with ISGRI \citep{DeBecker+07}.

Assuming that all the synchrotron emission produced in the shock region could escape without suffering a significant absorption, an upper limit of the magnetic field in the shock region can be estimated: $B^2 < 8\pi U_{ph} L_{radio}/L_{IC}$, where $U_{ph} \sim 0.1$ is derived from the luminosity and orbital size of \object{$\eta$~Car} \citep[see e.g.][]{Kashi+07}. The observed radio flux at 3.5 cm is typically around 5\% of the IC emission observed by \textit{INTEGRAL} \citep{Kashi+07}. Although it is likely that a substantial fraction of the synchrotron flux is actually absorbed in the wind interaction zone, it is remarkable that the derived value of $\sim 0.3 G$ is comparable with the values inferred for other CWB systems \citep[see e.g.][]{Benaglia+03}.

The IC emission of \object{$\eta$~Car} is expected to vary along the orbit, ranging from a weak emission with a low-energy cutoff around periastron, to a detectable flux and a cutoff in the GeV region when the soft photon density decreases \citep{Reimer+06}. The spectral slope integrated over the shock region is believed to vary between 1.5 and 2, depending on the geometry of the system.

The closest EGRET source is \object{3EG~1048-5840} \citep{Hartman+99}, located 1.1\degree\ away from \object{$\eta$~Car}, and which was associated with \object{PSR~J1048-5832} \citep{Kaspi+00}. Even when using this EGRET spectrum as an upper limit for the average GeV emission of \object{$\eta$~Car}, the spectral slope observed with ISGRI is significantly harder than a powerlaw extrapolation between the ISGRI measurements and the EGRET upper limits, thus indicating that the high-energy spectrum of \object{$\eta$~Car} should gradually steepen with energy in the 100~MeV--1~GeV region. This is in reasonable agreement with the predictions of \citet{Reimer+06}, and makes \object{$\eta$~Car} a source likely to be detected by \textit{Agile}, \textit{GLAST}, and perhaps \textit{HESS-II}.

\section{Conclusions}
\label{sec:Conclusions}
The first unambiguous detection of \object{$\eta$~Car} in the hard X-rays unveils a luminosity of $7 \times 10^{33}$~erg\,s$^{-1}$ (22--100~keV), i.e., 0.1\% of the kinetic energy available in the wind collision. This is the first observation of inverse Compton emission and particle acceleration in a colliding-wind binary. 
The absence of non-thermal radio emission allows to constrain the magnetic field in the particle acceleration region to be below 0.3~G. \object{$\eta$~Car} is expected to be detected in the GeV range, with \textit{Agile} and \textit{GLAST}. A detection (or an upper limit) with \textit{HESS-II} will also be very useful to constrain the emission spectrum.

Two other nearby hard X-ray sources have also been detected, the AXP \object{1E 1048.1-5937} and a new source (\object{IGR~J10447-6027}), which coincides with a YSO in the South of the Carina giant dust pillar nebula.

\begin{acknowledgements}
This research has made use of the SIMBAD database (CDS), of public data from \textit{INTEGRAL} (ESA) \& \textit{BeppoSAX} (ASI).
JCL and GR acknowledge support through the \textit{XMM-INTEGRAL} PRODEX project.
\end{acknowledgements}

\bibliographystyle{aa}
\bibliography{Article-Final}

\newpage

\end{document}